\documentclass[aps,prb,preprint,showpacs,groupaddress,floatfix]{revtex4-1}
\usepackage{graphicx,graphics}
\usepackage{dcolumn}
\usepackage{amsmath,amssymb,amsfonts}
\usepackage{latexsym,verbatim}
\usepackage{bm}
\usepackage{color}
\usepackage{ulem}
\usepackage{graphicx,epstopdf}
\usepackage[breaklinks=true,colorlinks,citecolor=blue,linkcolor=blue,urlcolor=blue]{hyperref}

\def\bef{\begin{framed}}
\def\eef{\end{framed}}
\def\be{\begin{equation}}
\def\ee{\end{equation}}
\def\ber{\begin{eqnarray}}
\def\eer{\end{eqnarray}}
\def\xiv{{\boldmath{\xi}}}

\def\xibold{\mbox{\boldmath $\xi$}}

\def\rv{{\bf r}}

\def\pv{{\bf p}}
\def\dv{{\bf d}}

\def\xv{{\bf x}}

\def\kv{{\bf k}}
\def\qv{{\bf q}}

\def\Gv{{\bf G}}

\def\Rv{{\bf R}}

\def\nn{\nonumber}
\def\xiv{\bm{\xi}}

\begin{document}
\title{Ab initio electronic structure of quasi two-dimensional materials: a ``native" gaussian--plane wave approach}
\author{Paolo E. Trevisanutto}
\affiliation{Graphene Research Centre and CA2DM, National University of Singapore, Singapore 117542, Singapore}
\affiliation{Singapore Synchrotron Light Source, National University of Singapore, Singapore 117603, Singapore}
\author{Giovanni Vignale}
\email{vignaleg@missouri.edu}
\affiliation{Department of Physics and Astronomy, University of Missouri, Columbia, Missouri 65211, USA}
\begin{abstract}
Ab initio electronic structure calculations of two-dimensional layered structures are typically performed using codes that were developed for three-dimensional structures, which are periodic in all three directions.  The introduction of a periodicity in the third direction (perpendicular to the layer) is completely artificial and may lead in some cases to spurious results and to difficulties in treating the action of external fields.  In this paper we develop a new approach, which is ``native" to quasi-2D materials, making use of basis function that are periodic in the plane, but atomic-like in the perpendicular direction.  We show how some of the basic tools of ab initio electronic structure theory -- density functional theory, GW approximation and Bethe-Salpeter equation -- are implemented in the new basis.  We argue that the new approach will be preferable to the conventional one in treating the peculiarities of layered materials, including the long range of the unscreened Coulomb interaction in insulators, and the effects of strain, corrugations, and external fields. 
\end{abstract}

\maketitle

\section{Introduction}
Following the successful isolation of single-layer graphene in 2004 \cite{Novoselov},  the study of quasi two-dimensional materials -- either single atomic layers or stacks of a few layers -- has become one of the most important areas of research in materials science.   It is expected that the unprecedented flexibility of these  materials, which can be assembled in many different combinations, including insulating, metallic, and superconducting layers, will lead in time to major technological advances.  Basic physics concepts such as massless Dirac Fermions and coupled valley-spin dynamics can also be studied in these systems.  From the point of view of computational electronic structure theory the challenge is to reliably predict the properties of a quasi-2D structure with minimal input from experiment.   Ideally we would like to be able to decide, based  on computation, which structures are stable, which are metals and which are insulators, what their excitation spectra look like.   Most of the methods  that are presently in use in solid state physics to address these questions --  Kohn-Sham density functional theory, GW approximations, Bethe-Salpeter (BS) equations, dynamical mean field theory --  have been developed for 3D systems and are now being applied to quasi-2D systems (important exceptions are Octopus\cite{Octopus} and GPAW \cite{gpaw}, which allow the user to specify a reduced number of periodic directions).
  
However, the transition from 3D to quasi-2D is not free of pitfalls.  Bulk 3D crystals exhibit periodicity in all three directions and are treated with the help of periodic boundary conditions.   This means, in practice, that computational methods are implemented on a basis of wave functions that exhibit three-dimensional periodicity, such as plane waves~\cite{Gonze20092582,QE-2009,gpaw,PhysRevB.54.11169}, augmented plane waves (APW) \cite{Exciting!,elk-code,PhysRevB.81.125102}, LMTO \cite{LMsuite} or periodic sums of atomic orbitals \cite{Fhi-Aims,Siesta,gpaw, Octopus}.  Quasi 2D materials on the other hand,  have crystalline periodicity in two dimensions only.  In the third direction (which we will call the $z$-direction)  the wave function is that of a polyatomic molecule, possibly a very large one, but in any case decaying exponentially at sufficiently large distance from the atomic layers. Therefore, strictly speaking, periodic boundary conditions cannot be used in the third direction.  Another key difference is that, due to the quasi-two dimensional character of the electronic charge distribution,  electric field fluctuations are not screened at sufficiently large distance in the plane of the layers \cite{PhysRevB.84.085406,PhysRevB.92.245123}: therefore electron-electron interaction effects are much stronger than in conventional 3D materials.  

It turns out that these difficulties are connected.  In order to keep using 3D computer codes with periodic basis sets it is customary to resort to an artifice known as the  ``supercell method", whereby the quasi-2D material is periodically repeated in the $z$ direction.  The distance between the artificial replicas, $d$,  is chosen to be large in the hope that interaction effects between the replicas can be ignored.  If everything goes well, the results should  become independent of  $d$ in the limit $d \to \infty$.  Unfortunately, because of the failure of screening, the Coulomb interaction between artificial replicas decreases very slowly with increasing $d$, and may cause spurious results.  In addition the exchange-correlation potentials of density functional theory, when treated exactly, are known to be highly nonlocal functionals of the density, which decay very slowly, with the inverse of the distance from the plane of the material \cite{Lang_Kohn,vBarth,Silkin}.  The problem is particularly acute in the calculation of dynamical response properties, where the replicas create an artificial dynamical environment.  

To counter these difficulties two main techniques have been developed so far.  The first ~\cite{Rozzi_cutoff,Castro_cutoff} relies on truncating the Coulomb interaction beyond a certain cutoff distance {in the $z$ direction}. Thus one hopes  eliminate  the unphysical interaction between the replicas while preserving the long-range interaction within the physical layer, or set of layers.   Another solution~\cite{Nazarov_2d} is to calculate the mean field created by the replicas and subtract it from the field that acts on the electrons in the physical layer: for the dynamical dielectric function and related properties this can be done in a relatively inexpensive way. 

Ultimately,  the best and most elegant solution would be to do the calculations with  a code that recognizes at the outset the quasi-2D character of the system and takes advantage of its structure by using a basis set of wave functions that are periodic (plane-wave like) in the $(x,y)$ plane and atomic-like in the $z$-direction.  We refer to such a code as a ``native" quasi-2D code.  A similar approach has been proposed, very recently,  in Ref.~[\onlinecite{PhysRevB.92.115132}], where the full-potential linearized augmented plane-wave method is modified to account for the exponential decay of the basis function in the $z$-direction. 

In this paper we take a first step towards the construction of a native quasi-2D code by introducing a set of  basis functions with the properties outlined above,  and by showing  how some of the basic tools of ab initio electronic structure theory -- density functional theory, GW approximation and Bethe-Salpeter equation -- are implemented in this basis.  Our approach is inspired by the FHI-AIMS basis set, in which periodic functions in 3D are constructed from Bloch sums of numerically generated  {Natural Atom-centred Orbitals (NAO) - see Appendix \ref{app_c}.  We insert these orbitals into two-dimensional Bloch sums to arrive at the desired basis functions.  While the FHI-AIMS code is already able to perform ``native" Kohn-Sham DFT on a quasi-2D basis,   our implementation of GW and BS equations on this basis set is new and will prove to be a powerful and versatile tool for the calculation of electronic properties of quasi-2D materials.  

It might be objected that the use of NAO orbitals for the $z$-dependence of the wave function will become unpractical (i.e., require many terms for good convergence)  when the wave function is very extended in that direction, as is the case for high-energy unoccupied states.  This is partially true, but we counter that (i) these extended states would become anyway unreliable when treated in the conventional supercell method, due to spurious cross-talk between replicas, (ii) extended basis sets have recently been developed and optimized to treat the excited states in FHI-AIMS\cite{Extended_aims1,Extended_aims2}, and  (iii) a new technology exists, based on the Sternheimer equation \cite{PhysRevB.81.115105,PhysRevB.88.075117} to avoid the use of excited states in the implementation of the GW method.  
Finally, we expect that the two-dimensional implementation of these tasks will speed up the calculations and make them more reliable, once the required computer codes are fully developed.

This paper is organized as follows: in section \ref{section_repre_vectors} we introduce the quasi-2D basis functions and review their formal properties.
In particular, because these functions are not mutually orthogonal, we find it convenient to introduce covariant and contravariant components of vectors in the Hilbert space -- the two types of components being connected by a ``metric tensor", which is the matrix of the overlaps. In section \ref{section_repre_two_points}, the two space point functions are described in this new formalism. Therefore, we represent the Dyson equation for the GW self-energy (section \ref{section_gw}), the polarization function (section \ref{section_pol_function}) and screened potential W (section \ref{section_screen_inter}) on the quasi-2D basis.  Finally, in section \ref{section_bse}, the BS equation for the two-particle propagator on the quasi-2D basis is made explicit and we introduce the effective two-band exciton equation (sub section \ref{subsection_bse}).  The last section \ref{summary} contains our summary and outlook for the development of the ideas presented in this paper. In appendix \ref{app_a} a  single pole approximation is revisited. In appendix \ref{app_b}, the contour integration technique for the GWA is derived in our formalism. Finally,  a brief overview of the FHI-AIMS NAO orbitals is provided in appendix \ref{app_c}.

\section{Hybrid basis set and representation of vectors}
\label{section_repre_vectors}
For a system that is periodic in the $(x,y)$ plane, but not in the $z$ direction --  an array of molecules arranged in a 2D lattice -- we introduce basis functions of the Bloch type, characterized by a Bloch wave vector $\kv$ in the $(x,y)$ plane.  The general form of these basis functions is~\cite{PhysRevB.48.17791,PhysRevLett.75.3489,PhysRevB.62.4927} 
\be\label{CovariantBasis}
\chi_{\alpha\kv}(\xibold)\equiv\frac{1}{\sqrt{N}}\sum_{\Rv} e^{i \kv \cdot\Rv}\phi_{nlm}(\xibold -\Rv-\dv_a)
\ee 
where  $\xibold = (\xv,z)$ is a three dimensional position vector; $\Rv$ is a lattice vector in the $(x,y)$  plane, $\dv_a$ is a three-dimensional vector specifying the position of the $a$-th atom in the two-dimensional unit cell labelled by $\Rv$, and $\phi_{nlm}(\xibold -\Rv-\dv_a)$ are orthonormal atomic orbitals labeled by atomic quantum numbers $(n,l,m)$ and centered at $\Rv+\dv_a$.   $N$ is the number of cells in the two-dimensional lattice, and $\alpha = n,l,m,a$ is a composite index including the atomic quantum numbers and the position of the atom in the unit cell. The form of the $\phi$ orbitals will be left unspecified for the time being: explicit numerical expressions that are used in the FHI-AIMS code will be discussed later (see  Appendix \ref{app_b})}.   Note that
\be
\chi_{\alpha\kv}(\xibold+\Rv)=e^{i \kv \cdot\Rv} \chi_{\alpha\kv}(\xibold)\,,
\ee
as required for a Bloch wave of wave vector $\kv$.

The basis functions satisfy the ``orthonormality" relation
\be
 \int d\xibold  \chi^{*}_{\alpha\kv}(\xibold)\chi_{\beta \pv}(\xibold)  \equiv S_{\alpha\beta}(\kv)\delta_{\kv,\pv}\,,
 \ee
where $S_{\alpha\beta}(\kv)=S^*_{\beta\alpha}(\kv)$ is a hermitian matrix.
In addition to the covariant basis functions of Eq.~(\ref{CovariantBasis}) it is convenient to introduce the contravariant basis set
\be
\chi^{\alpha}_{\kv}(\xibold)  \equiv  \sum _{\beta} S^{\alpha \beta}(\kv) \chi_{\beta\kv}(\xibold)
\ee
where $S^{\alpha\beta}(\kv)\equiv[S^{-1}(\kv)]_{\beta\alpha}$ is the transposed inverse of the ``metric tensor" $S(\kv)$. In Quantum Chemistry, $S^{\alpha\beta}$ is referred as overlap matrix. In terms of covariant and contravariant basis elements the ``orthonormality" relations take the simpler form
\be\label{Orthonormality2}
\int d\xibold  \chi^{\alpha*}_{\kv}(\xibold)\chi_{\beta \pv}(\xibold)  =\delta_{\alpha\beta}(\kv)\delta_{\kv,\pv}\,.
 \ee
 We can also write the completeness relation in the form
 \be\label{Completeness}
 \sum_{\alpha\kv}\chi_{\alpha \kv}(\xibold)\chi^{\alpha*}_{\kv}(\xibold') = \delta(\xibold-\xibold')\,,
 \ee
 where the sum over $\kv$ runs over the vectors of the first Brillouin zone of the reciprocal 2D lattice.  This is evident from the fact that for an arbitrary function  $f(\xiv)=\sum_{\beta \pv} f^\beta_{\pv}\chi_{\beta\pv}(\xiv)$ we have,  by virtue of Eq.~(\ref{Orthonormality2}),
 \be
 \int d\xiv' \sum_{\alpha\kv}\chi_{\alpha \kv}(\xibold)\chi^{\alpha*}_{\kv}(\xibold') f(\xiv') = \sum_{\alpha\kv} f^\alpha_{\kv}\chi_{\alpha\kv}(\xiv)  = f(\xiv)\,.
 \ee

A Bloch wave function $\psi_{\kv}(\xibold)$ in the Hilbert space can be expressed as a contravariant vector as follows:
\be
\psi_\kv(\xibold) = \sum_{\alpha}\psi_{\kv}^\alpha \chi_{\alpha\kv}(\xibold)\,,~~~{\rm where}~~~ \psi_{\kv}^\alpha =\int d\xibold \chi^{\alpha*}_{\kv}(\xibold) \psi_{\kv}(\xibold)\,.
\ee
%or, alternatively,  as a covariant vector
%\be
%\psi_\kv(\xibold) = \sum_{\alpha}\psi_{\kv\alpha} \chi^{\alpha}_{\kv}(\xibold)\,,~~~{\rm where}~~~ \psi_{\kv\alpha} =\int d\xibold \chi^{*}_{\alpha\kv}(\xibold) \psi_{\kv}(\xibold)\,.
%\ee
The scalar product of two Bloch wave functions $\psi_{\kv}(\xibold)$ and $\varphi_{\kv}(\xibold)$ is then given by
\be
(\psi_{\kv},\varphi_{\kv}) = \int d\xibold\psi^*_{\kv}(\xibold)\varphi_{\kv}(\xibold) = \sum_\alpha \psi^*_{\kv\alpha} \varphi_{\kv}^{\alpha} = \sum_\alpha \psi^{*\alpha}_{\kv} \varphi_{\kv\alpha} \,. 
\label{Bloch_wf}
\ee

Covariant and contravariant components of the same vector are  connected by the metric tensor:
\be
\psi_{\kv\alpha} = \sum_{\beta}\psi_{\kv}^{\beta}S_{\beta\alpha}(\kv) \,.
\ee

In the following, the solutions of, say, the Kohn-Sham equations for a given layered structure will be represented as covariant or contravariant vectors in the hybrid atomic-plane wave basis.

 \section{Representation of two-point functions}
 \label{section_repre_two_points}
 We now consider the representation of a periodic two-point function $f(\xibold,\xibold')$, i.e., a function such that 
 \be
 f(\xibold+\Rv,\xibold'+\Rv)=f(\xibold,\xibold')\,,
 \ee
 where $\Rv$ is a two-dimensional lattice vector.  Making use of the completeness relation~(\ref{Completeness}) we easily see that
 \be\label{Eq12}
 f(\xibold,\xibold')=\sum_{\alpha\beta\kv} \chi_{\alpha\kv}(\xibold) f^{\alpha\beta}(\kv) \chi_{\beta\kv}^*(\xibold')\,,~~~~{\rm where}~~~f^{\alpha\beta}(\kv)=\int d\xibold d\xibold'  \chi^{*\alpha}_{\kv}(\xibold)f(\xibold,\xibold')
 \chi^{\beta}_{\kv}(\xibold')
 \ee
 
 or, alternatively
 \be\label{Eq13}
 f(\xibold,\xibold')=\sum_{\alpha\beta\kv} \chi^{\alpha}_{\kv}(\xibold) f_{\alpha\beta}(\kv) \chi^{*\beta}_{\kv}(\xibold')\,,~~~~{\rm where}~~~f_{\alpha\beta}(\kv)=\int d\xibold d\xibold'  \chi^{*}_{\alpha\kv}(\xibold)f(\xibold,\xibold')
 \chi_{\beta\kv}(\xibold')
 \ee
  
 We notice in passing that the first of Eqs.~(\ref{Eq12}) and the second of Eqs.~(\ref{Eq13}) correspond, respectively, to Eqs. (6) and (7) of Ref.~\onlinecite{PhysRevB.48.17791}, i.e., $f^{\alpha\beta}$ and $f_{\alpha\beta}$ correspond, respectively, to the ``angular bracket" and ``square bracket" matrix elements of Ref.~\onlinecite{PhysRevB.48.17791}.
  
 These representations apply, in particular to the Green's function $G(\xibold,\xibold',\omega)$ and the screened electron-electron interaction $W(\xibold,\xibold',\omega)$.  We now show how the two representations are combined in the GW approximation.

As an example, consider the noninteracting Green's function (advanced and retarded Green's function):
\begin{equation}\label{GF1}
  G_0(\xibold,\xibold',\omega)= \sum_{n{\bf k}}\frac{\psi_{n\kv}(\xibold)\psi_{n\kv}^{*}(\xibold')} {\omega- E_{n\kv}\pm i0^+ }\,.
\end{equation}

 Then, expanding in Bloch  functions:
 \be\label{GF2}
 G_0^{\alpha\beta}(\kv,\omega)= \sum_{n}\frac{\psi^{\alpha}_{n{\kv}}\psi^{\beta*}_{n\kv}} {\omega-E_{n\kv}\pm i0^+}
 \ee
 where $\psi_{n\kv}(\xiv)\equiv\sum_{\alpha}\psi^{\alpha}_{n{\kv}}\chi_{\alpha\kv}(\xibold)$ are non-interacting one- electron wave functions and $E_{n\kv}$ the one-electron energies.

\section{GW Self-Energy}\label{section_gw}
\begin{figure}
\includegraphics[width=4.0 in]{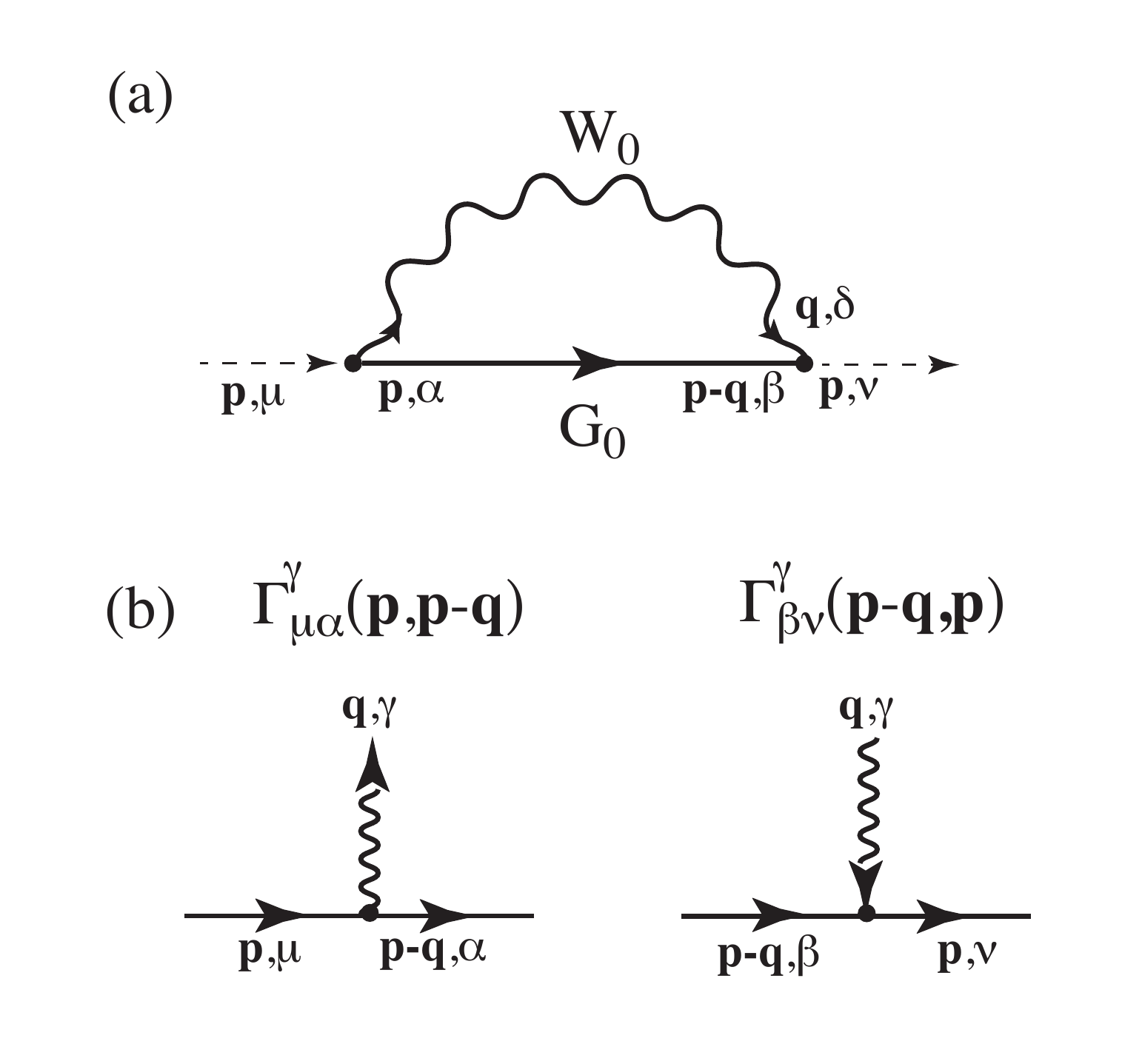}
\caption{Diagrammatic representation of the $G_0W_0$ self-energy and the three-point vertices.}
\label{GW}
\end{figure}

In the GW approximation the self-energy is a parallel convolution product of G and W, usually approximated by their lowest-order forms $G_0$ and $W_0$:
\be
\Sigma(\xibold,\xibold',\omega)=\frac{i}{2\pi}G_0(\xibold,\xibold',\omega)*W_0(\xibold,\xibold',\omega)e^{-i\eta\omega}\,.
\ee
Here the $*$ represent the convolution product: $f(\omega)*g(\omega)\equiv \int d\omega' f(\omega-\omega')g(\omega')$ and .
The corresponding diagram is shown in Fig.~(\ref{GW}).

We expand the product as
\be
\Sigma(\xibold,\xibold',\omega)= \sum_{\pv'\qv}G_{0}^{\alpha\beta}(\pv',\omega)\chi_{\alpha\pv'}(\xibold)\chi^*_{\beta\pv'}(\xibold')*W_{0,\gamma\delta}(\qv,\omega)\chi^{\gamma}_{\qv}(\xibold)\chi^{*\delta}_{\qv}(\xibold')
\ee
where the sum over repeated indices is understood. 
Then the matrix element of the self-energy is
\be
\Sigma_{\mu\nu}(\pv,\omega) =\frac{i}{2\pi} \sum_{\pv'\qv} \left[\int d\xibold \chi^*_{\mu\pv}(\xibold)\chi_{\alpha\pv'}(\xibold) \chi^{\gamma}_{\qv}(\xibold)\right]G_{0}^{\alpha\beta}(\pv',\omega) \left[\int d\xibold' \chi^*_{\beta\pv'}(\xibold') \chi^{*\delta} _{\qv}(\xibold')  \chi_{\nu\pv}(\xibold') \right]*W_{0,\gamma\delta}(\qv,\omega)
\ee
The quantities in the square brackets vanish unless $\pv'=\pv-\qv$.  We define
\be \label{Gamma1}
\int d\xibold \chi^*_{\mu\pv}(\xibold)\chi_{\alpha\pv'}(\xibold) \chi^{\gamma}_{\qv}(\xibold) \equiv \Gamma^{\gamma}_{\mu\alpha}(\pv,\pv-\qv)
\ee
and
\be \label{Gamma2}
\int d\xibold' \chi^*_{\beta\pv'}(\xibold') \chi^{*\delta} _{\qv}(\xibold')  \chi_{\nu\pv}(\xibold') \equiv \Gamma^{\delta}_{\beta\nu}(\pv-\qv,\pv)\,.
\ee
These matrix elements are illustrated in Fig.~(\ref{GW} (b)).  Note that $\Gamma^{\delta}_{\beta\nu}(\pv-\qv,\pv) = [\Gamma^{\delta}_{\nu\beta}(\pv,\pv-\qv)]^*$.
Then complete result for the self-energy is
\be
\Sigma_{\mu\nu}(\pv,\omega) =\frac{i}{2\pi}\sum_{\qv}\Gamma^{\gamma}_{\mu\alpha}(\pv,\pv-\qv)G_{0}^{\alpha\beta}(\pv-\qv,\omega)\Gamma^{\delta}_{\beta\nu}(\pv-\qv,\pv) *W_{0,\gamma\delta}(\qv,\omega)
\ee

\section{Polarization function}\label{section_pol_function}
\begin{figure}
\includegraphics[width=4.0 in]{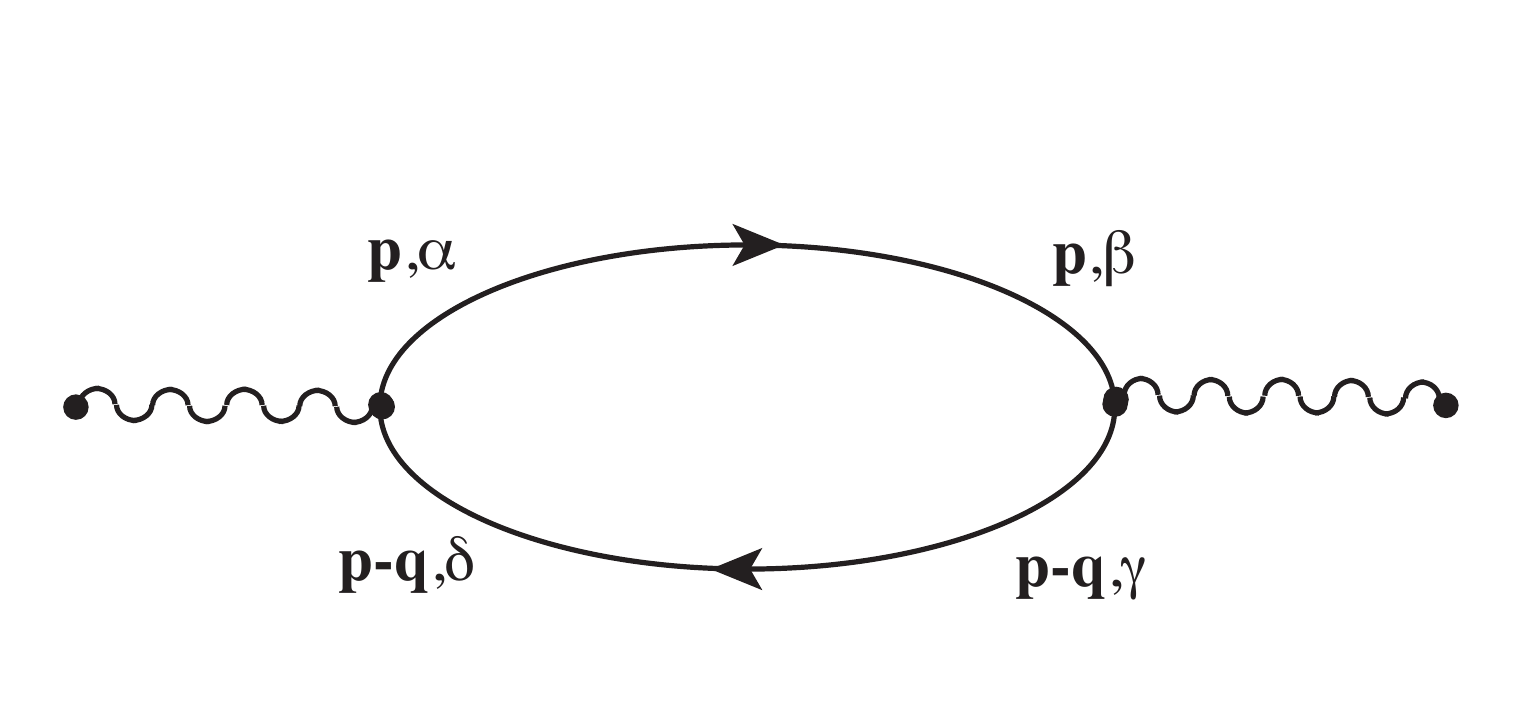}
\caption{Diagrammatic representation of the polarization propagator.}
\label{P0}
\end{figure}

Another crucial two-point function is the polarization function.  The noninteracting polarization function is an antiparallel convolution of two Green's functions
\be
P_0(\xibold,\xibold',\omega)=G_{0}(\xibold,\xibold',\omega)*G_{0}(\xibold',\xibold,-\omega)\,,
\ee
where $*$ represents the convolution product. A graphical representation of this formula is shown in Fig.~\ref{P0}.

We expand the product as
\be
P_0(\xibold,\xibold',\omega)= \sum_{\pv\pv'}G_{0}^{\alpha\beta}(\pv,\omega)\chi_{\alpha\pv}(\xibold)\chi^*_{\beta\pv}(\xibold')*G_{0}^{\gamma\delta}(\pv',-\omega)\chi_{\gamma\pv'}(\xibold')\chi^*_{\delta\pv'}(\xibold)
\ee
Then the matrix element of the polarization propagator is
\be
P_0^{\mu\nu}(\qv,\omega) =\sum_{\pv\pv'}\left[\int d\xibold \chi^{*\mu}_\qv (\xibold) \chi_{\alpha\pv}(\xibold)  \chi^*_{\delta\pv'}(\xibold) \right] G_{0}^{\alpha\beta}(\pv,\omega)\left[\int d\xibold'\chi^*_{\beta\pv}(\xibold') \chi^{\nu}_{\qv}(\xibold')  \chi_{\gamma\pv'}(\xibold')\right]*G_{0}^{\gamma\delta}(\pv',-\omega)
\label{Polarization_af}
\ee
Again,  the quantities in the square brackets vanish unless $\pv'=\pv-\qv$.  With the three-point vertices $\Gamma$ defined in Eqs.~(\ref{Gamma1}) and ~(\ref{Gamma2}) we easily get the following representation of the polarization propagator: 
%\be
%\int d\xibold \chi^{*\mu}_\qv (\xibold) \chi_{\alpha\pv}(\xibold)  \chi^*_{\delta\pv'}(\xibold) =  \Gamma^\mu_{\delta\alpha}(\pv-\qv,\pv)
%\ee
%and
%\be
%\int d\xibold'\chi^*_{\beta\pv}(\xibold') \chi^{\nu}_{\qv}(\xibold')  \chi_{\gamma\pv'}(\xibold') = \Gamma^\nu_{\beta\gamma}(\pv,\pv-\qv)\,.
%\ee
%Thus we have
\be
P_0^{\mu\nu}(\qv,\omega) =\sum_{\pv}\Gamma^\mu_{\delta\alpha}(\pv-\qv,\pv)
 G_{0}^{\alpha\beta}(\pv,\omega)\Gamma^\nu_{\beta\gamma}(\pv,\pv-\qv)*G_{0}^{\gamma\delta}(\pv-\qv,-\omega)
\ee

\section{The screened interaction}\label{section_screen_inter}
The covariant  components of the RPA screened interaction can be obtained from the formula
\be
W_{0,\alpha\beta}(\qv,\omega) = [v^{1/2}(\qv)]_{\alpha\gamma}[\tilde \epsilon^{-1}(\qv,\omega)]^{\gamma\delta}   [v^{1/2}(\qv)]_{\delta\beta}\,. 
\ee
Here 
\be
[v^{1/2}(\qv)]_{\alpha\gamma} = \int d\xibold d\xibold'  \chi^*_{\alpha\qv}(\xibold)v^{1/2}(\xibold,\xibold')\chi_{\gamma\qv}(\xibold')
\ee
where
\be
v^{1/2}(\xibold,\xibold') = \frac{e}{\pi^{3/2}|\xibold-\xibold'|^2}\,,
\ee
is the ```square root" of the Coulomb interaction.  Further we have
\be
\tilde \epsilon_{\alpha\beta}(\qv,\omega) = S_{\alpha\beta}(\qv) - [v^{1/2}(\qv)]_{\alpha\gamma}  P_0^{\gamma\delta}(\qv,\omega)[v^{1/2}(\qv)]_{\delta\beta}
\ee

A simple single-pole approximation for $\tilde \epsilon(\qv,\omega)$ is described in the Appendix.

 \section{The Bethe-Salpeter Equation}\label{section_bse}
 
\begin{figure}
\includegraphics[width=5.0 in]{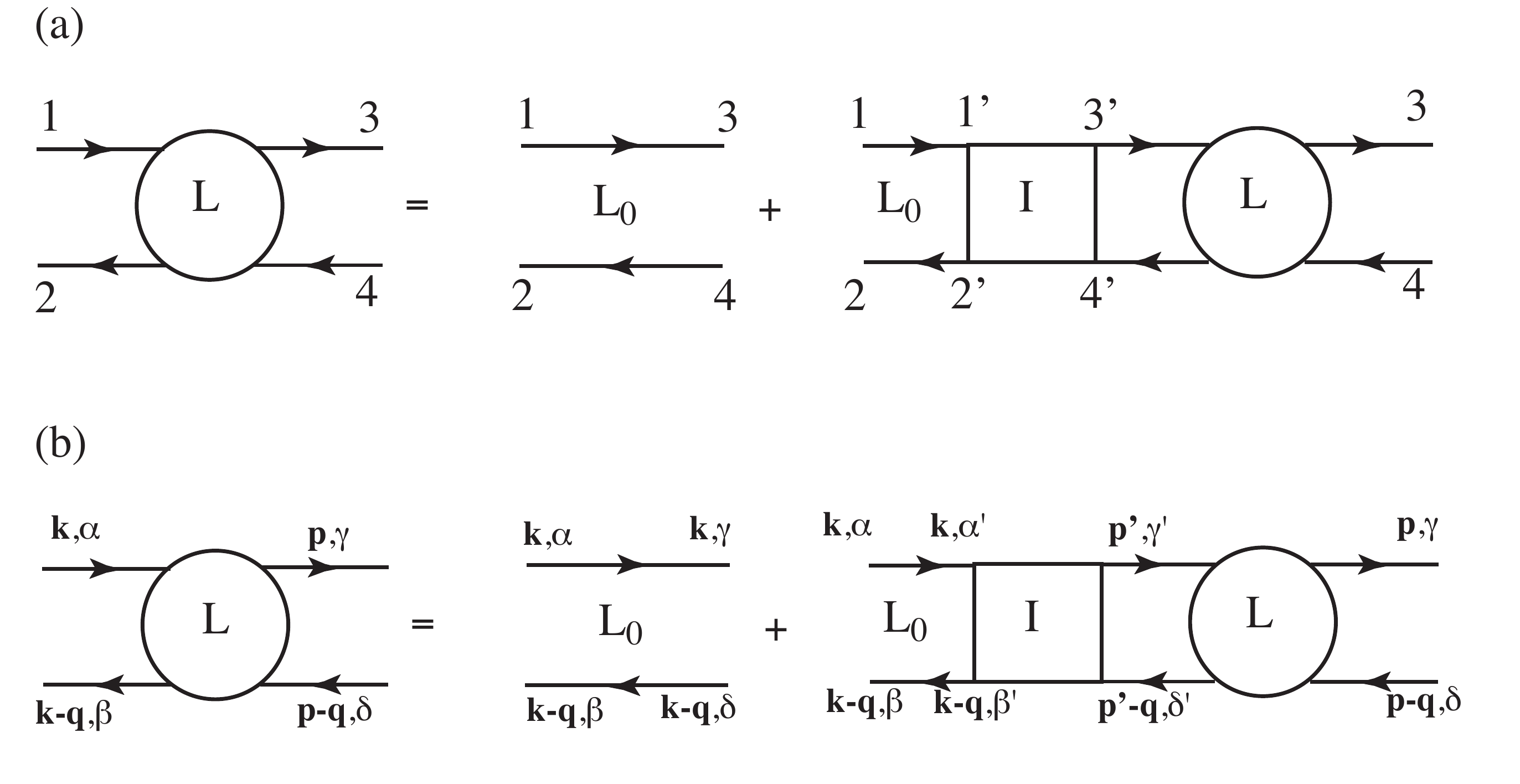}
\caption{Diagrammatic representation of the Bethe-Salpeter equation in real space (a) and in the localized basis (b).}
\label{BSE}
\end{figure}

Besides one-particle properties,  which are usually described in terms of the Kohn-Sham wave functions and the self-energy, two-particle properties also play a prominent role in computational materials science.  For example, the information about the optical spectrum and the refractive index of materials is contained in the macroscopic dielectric function $\epsilon_M(\omega)$, which, in ordinary three-dimensional periodic systems, is defined as the ratio between the  macroscopic external electric field and the macroscopic total electric field, where ``macroscopic" means averaged over a region much larger than the size of the unit cell.  For 2D materials care must be exerted to construct the appropriate response function for the experimental situation under study\cite{Nazarov_2d}.  In the case of optical properties, the macroscopic dielectric function reduces to $1$ and we focus instead on the longitudinal in-plane polarizability per unit cell, which we define as
\be\label{P00}
\chi_P(\omega)= \frac{1}{N_c}\lim_{\qv \to 0}\frac{e^2}{q^2 }\int d\xiv \int d\xiv' e^{-i\qv\cdot(\rv-\rv')} \bar L(\xiv,\xiv,\xiv',\xiv';\omega)\,,
\ee
%The averaging over the thickness of the slab, however,  requires that we introduce a suitable weight function $F(z)$, for example the normalized density profile of the material, averaged over the two-dimensional unit cell, in such a way that $\int F(z) dz =1$ and $F(z)$ is exponentially localized about $z=0$.  
%With this provision, we define
%\begin{equation}
%\epsilon_{M}(\omega)= \frac{1}{AL}\left( \int d \xiv  \int d\xiv' \epsilon^{-1}(\xiv,\xiv',\omega) \right)^{-1}\,,
%\label{Macroeps2}
%\end{equation}
%where  the integral is performed over the area $A$ the material and a length $L$much larger than the thickness of the slab in the $z$ direction.  
%This is a dimensionless quantity, as follows from the fact that the microscopic dielectric function $\epsilon^{-1} (\xiv,\xiv',\omega)$ has the dimensions of an inverse volume.   
%Further, it can be shown that the standard  treatment  of $\epsilon_M(\omega)$ in three-dimensional periodic systems  (see Appendix B of Ref. \cite{rmp_onida}) can be generalized to the present case  (see discussion in Appendix D), yielding 
%\begin{equation}
%\epsilon_{M}(\omega) =  1 - \lim_{\qv \to 0} \frac{2\pi e^2}{q}  \bar P_{00}(\qv,\omega)
%\label{Macroeps}
%\end{equation}
% where 
%\be\label{P00}
%\bar P_{00}(\qv,\omega) = \int d\xiv \int d\xiv' e^{-i\qv\cdot(\rv-\rv')} \bar L(\xiv,\xiv,\xiv',\xiv';\omega)\,.
%\ee
where $\bar L(\xiv_1,\xiv_2,\xiv_3,\xiv_4;\omega)$ is the proper four-point electron-hole propagator, integrated over the frequencies of the external legs and $N_c$ is the number of unit cells.  We remind the reader that the most general form of the electron-hole propagator is $\bar L_{full}(\xiv_1,\xiv_2,\xiv_3,\xiv_4; \epsilon_1,\epsilon_2,\epsilon_3,\epsilon_4)$, where $\epsilon_2 = \epsilon_1-\omega$ and $\epsilon_4=\epsilon_3-\omega$ are the frequencies of the external legs.  The simpler propagator $L$ (still referred to as electron-hole propagator for brevity)  is obtained from $L_{full}$ by integrating over $\epsilon_1$ and $\epsilon_3$ at fixed $\omega$.  The bar over $L$ is a reminder that we are calculating $L$  {\it with exclusion of the long-range Coulomb interaction}~\cite{rmp_onida} in the so-called exchange electron-hole channel.
To explain this point more clearly we recall that the interacting electron-hole propagator is given by the solution of the Bethe-Salpeter Equation (BSE) (Fig. \ref{BSE} (a))
\be\label{BSEL}
\begin{split}
L(\xibold_1,\xibold_2,\xibold_3,\xibold_4;\omega) &= L_{0}(\xibold_1,\xibold_2,\xibold_3,\xibold_4;\omega)\\
 &+\int d\xibold'_1d\xibold'_2d\xibold'_3d\xibold'_4 L_{0}(\xibold_1,\xibold_2,\xibold_1',\xibold_2';\omega)I(\xibold'_1,\xibold_2',\xibold_3',\xibold_4')L(\xibold_3',\xibold_4',\xibold_3,\xibold_4;\omega)
\end{split}
\ee
where $L_{0}$ is the non interacting electron-hole propagator (also integrated over the frequencies of the external legs)  and $I$ is the {\it irreducible} 4-point interaction function, which includes only diagrams that cannot be separated into two parts by cutting a single  noninteracting electron-hole propagator $L_{0}$ (see Fig.~\ref{BSE} (b)).   Notice that in writing the above equation we have assumed that the irreducible 4-point interaction is instantaneous in time, i.e., it does not depend on the frequencies of the external legs.  This is, of course, an uncontrolled approximation, and we know that it misses some important physics\cite{Romaniello,Strinati}   -- yet it is this assumption that enables us to write a closed equation for the frequency-integrated electron-hole propagators.

\begin{figure}
\includegraphics[width=5.0 in]{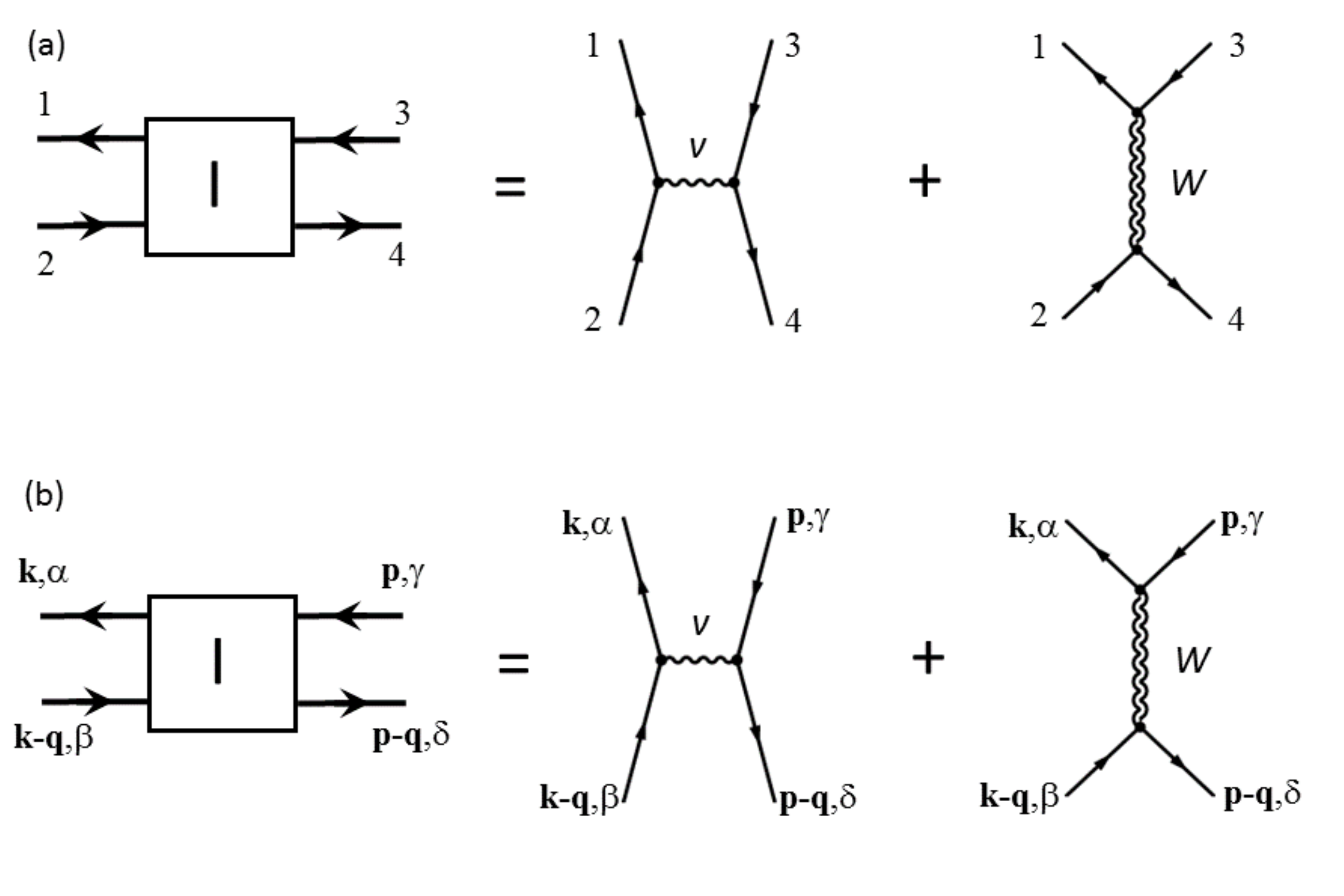}
\caption{Diagrammatic representation of the Bethe-Salpeter Kernel $I$ for the electron-hole Green's function in real space (a) and in the gaussian-plane wave basis (b).}
\label{KernelBSE}
\end{figure}
The instantaneous irreducible interaction $I$ contains, among many terms, an exchange-type electron-hole interaction of the form
\be
I^{x}(\xibold_1,\xibold_2,\xibold_3,\xibold_4)=-i\delta(\xibold_1-\xibold_2)\delta(\xibold_3-\xibold_4)v(\xibold_1-\xibold_3)\,
\label{kernel_hartree}
\end{equation}
(see Fig.~(\ref{KernelBSE}(a))).
The  Coulomb interaction can be separated into a long-range part
\be
v_{lr}(\xiv-\xiv') = \int_{BZ2} \frac{d^2 \qv}{(2\pi)^2}\frac{2\pi e^2}{q} e^{-q|z-z'|}e^{i\qv\cdot(\rv-\rv')}
\ee 
 where the integral runs over all values of $q_z$ and over $\qv$ in the first Brillouin zone of the two-dimensional periodic lattice (BZ2), and a short-range part
 \be
\bar v(\xiv-\xiv') = \sum_{\Gv \neq 0}\int_{BZ2} \frac{d^2 \qv}{(2\pi)^2}\frac{2\pi e^2}{|\qv+\Gv|} e^{-|\qv+\Gv||z-z'|}e^{i(\qv+\Gv)\cdot(\rv-\rv')}\,.
\ee
where  the sum runs over all the two-dimensional reciprocal lattice vectors different from {\bf 0}. 
The  electron hole propagator $\bar L$, which enters Eq.~(\ref{P00}), satisfies the same equation as $L$ (Eq.~(\ref{BSEL})), the only difference being that the long-range part of the interaction  is excluded: only  the short-range part, $\bar v$, is retained in $I^x$.
   
% Note that, because $\bar \qv$ is three-dimensional, while the $\Gv$'s are two-dimensional, what we are doing here is a slight generalization of the standard procedure of elimination of the long-range part of the Coulomb interaction, described in Appendix B of Ref.~\onlinecite{rmp_onida}.  The essential point is that, after excluding the $\Gv=0$ component of the Coulomb interaction, the additional exclusion of $q_z=0$ component is inconsequential, because it removes a finite integrand from a zero-dimensional region of phase space.
%

The BSE for $\bar L$ has the form
\be
\begin{split}
L(\xibold_1,\xibold_2,\xibold_3,\xibold_4;\omega) &= L_{0}(\xibold_1,\xibold_2,\xibold_3,\xibold_4;\omega)\\
 &+\int d\xibold'_1d\xibold'_2d\xibold'_3d\xibold'_4 L_{0}(\xibold_1,\xibold_2,\xibold_1',\xibold_2';\omega)\bar{I}(\xibold'_1,\xibold_2',\xibold_3',\xibold_4')\bar{L}(\xibold_3',\xibold_4',\xibold_3,\xibold_4;\omega)
\end{split}
\ee
where $\bar I$ is $I$ minus the long-range interaction term $I^{x}_{lr}(\xibold_1,\xibold_2,\xibold_3,\xibold_4)=-i\delta(\xibold_1-\xibold_2)\delta(\xibold_3-\xibold_4)v_{lr}(\xibold_1-\xibold_3)$.

A further approximation that is commonly made at this point is to approximate the irreducible interaction as the sum of short-range exchange and direct terms (Fig.~\ref{KernelBSE}(a)):
\be
\bar I(\xibold_1,\xibold_2,\xibold_3,\xibold_4)= \bar I^{x}(\xibold_1,\xibold_2,\xibold_3,\xibold_4)+\bar I^{d}(\xibold_1,\xibold_2,\xibold_3,\xibold_4)\,,
\ee
where
\be
\bar I^{x}(\xibold_1,\xibold_2,\xibold_3,\xibold_4)  = -i\delta(\xibold_1-\xibold_2)\delta(\xibold_3-\xibold_4)\bar v(\xibold_1-\xibold_3)
\ee
and
 \begin{equation}
    \bar I^{d} (\xibold_1,\xibold_2,\xibold_3,\xibold_4) = \delta(\xibold_1-\xibold_3)\delta(\xibold_2-\xibold_4)W_{0}(\xibold_1-\xibold_2, \omega=0)
    \label{kernel_corr}
    \end{equation}
where $W_{0}(\xibold_1-\xibold_2, \omega=0)$ represents the statically screened Coulomb interaction.

The four-point functions are straightforwardly represented in our basis set.  Taking as a prototype  $L(\xiv_1,\xiv_2,\xiv_3,\xiv_4;\omega)$  we see that this is replaced by the matrix $L_{\alpha\beta,\gamma\delta}(\kv,\kv',\qv;\omega)$ according to the scheme illustrated in the figure and analytically described by the equation
\be
L(\xiv_1,\xiv_2,\xiv_3,\xiv_4;\omega) = \sum_{\kv,\pv,\qv}\sum_{\alpha\beta\gamma\delta}\chi^\alpha_{\kv}(\xiv_1)[\chi^\beta_{\kv-\qv}(\xiv_2)]^*L_{\alpha\beta,\gamma\delta}(\kv,\pv,\qv;\omega)[\chi^\gamma_{\pv}(\xiv_3)]^*\chi^\delta_{\pv-\qv}(\xiv_4)
\label{L_real_space}
\ee
where
\be
L_{\alpha\beta,\gamma\delta}(\kv,\pv,\qv;\omega)=\int d\xiv_1\xiv_2\xiv_3\xiv_4 [\chi_{\alpha,\kv}(\xiv_1)]^*\chi_{\beta,\kv-\qv}(\xiv_2)
L(\xiv_1,\xiv_2,\xiv_3,\xiv_4;\omega)
\chi_{\gamma,\pv}(\xiv_3)[\chi_{\delta,\pv-\qv}(\xiv_4)]^*
\ee
In this basis, the BSE takes the form
\ber
L_{\alpha\beta,\gamma\delta}(\kv,\pv,\qv;\omega)&=&L_{0 (\alpha\beta,\gamma\delta)}(\kv,\pv,\qv;\omega)\nn\\&+& \sum_{\kv'\pv'}\sum_{\alpha'\beta'\gamma'\delta'}L_{0,(\alpha\beta,\alpha'\beta')}(\kv,\kv',\qv;\omega) I _{\alpha'\beta',\gamma'\delta'}(\kv',\pv',\qv)L_{\gamma'\delta',\gamma\delta}(\pv',\pv,\qv;\omega)\,.\nn\\
\eer
The noninteracting electron-hole propagator is given by
\be\label{non_inter_L0}
L_{0(\alpha\beta,\gamma\delta)}(\kv,\pv,\qv;\omega) = ~G_{0,\alpha\gamma}(\kv,\omega)* G_{0,\delta\beta}(\kv-\qv,-\omega) \delta_{\kv,\pv}\,.\
\ee
The frequency integral implied by the convolution product in this equation can be done analytically, with the help of the Kohn-Sham Green's functions given in Eq.~(\ref{GF2}), the result being
\be\label{L0abcd}
L_{0(\alpha\beta,\gamma\delta)}(\kv,\pv,\qv;\omega) =- i\delta_{\kv,\pv}\sum_{n,m}\psi_{\alpha,n\kv}\psi^*_{\gamma,n\kv}\frac{f(E_{n\kv})-f(E_{m\kv-\qv})}{\omega-E_{n\kv}+E_{m\kv-\qv}\pm i0^+}\psi_{\beta,m\kv}\psi^*_{\delta,m\kv}\,,
\ee
where $E_{n\kv}$ and $\psi^\alpha_{n\kv}$ are, respectively, the Kohn-Sham energy and the Kohn-Sham wave function of the Bloch state $n\kv$ in our representation,  and $f(E_{n\kv})$ is the Fermi-Dirac occupation factor of the same state.
% We do not give the explicit expression here, since it will be superceded in the next section by a simpler expression based on a ``single pole approximation" for the exciton (see Eq.~(\ref{PPEH})).
The  two components of the irreducible four-point interaction function are (Fig.~\ref{KernelBSE}(b))
\be
I^{x}_{\alpha\beta\gamma\delta}(\kv,\pv,\qv)=-i\sum_{\mu\nu}\Gamma^\mu_{\alpha\beta}(\kv,\kv-\qv)v_{\mu\nu}(\qv)\Gamma^\nu_{\delta\gamma}(\pv-\qv,\pv)
\ee
and
\be
I^{d}_{\alpha\beta\gamma\delta}(\kv,\pv,\qv)=\sum_{\mu\nu}\Gamma^\mu_{\alpha\gamma}(\kv,\pv)w_{\mu\nu}(\kv-\pv)\Gamma^\nu_{\delta\beta}(\kv-\qv,\pv-\qv)\,,
\ee
where the three-point vertices are defined by Eqs.~(\ref{Gamma1}) and ~(\ref{Gamma2}). Here $v_{\mu\nu}(\qv)$ denotes the matrix elements of the Coulomb interaction in our basis set, i.e.,
\be
v_{\mu\nu}(\qv)= \int d\xiv \int d\xiv'  \chi^*_{\mu\qv}(\xiv) v(\xiv-\xiv')\chi_{\nu\qv}(\xiv')
\ee
and similarly for $w_{\mu\nu}(\qv)$, where $w\equiv W_{0}(\omega=0)$ is the statically screened interaction in the same basis.  A similar expression holds for the regularized 4-point interaction $\bar I$ in terms of the regularized interaction $\bar v$, as previously discussed.

\subsection{Bethe-Salpeter equation in the single pole approximation }\label{subsection_bse}

The Bethe-Salpeter equation (BSE) for the electron-hole propagator is very complex, even after having made the simplifying assumption that the 4-point electron-hole interaction is instantaneous in time (i.e., the kernel $I$ has been taken to be frequency-independent).  A major simplification can be achieved if we restrict our attention to a single pair of bands, which we call $c$ for conduction and $v$ for valence and stipulate that the electron is in conduction band and the hole is in the valence band. Thus, in Eq.~(\ref{L0abcd}) we consider only the resonant term at the frequency $\omega = E_{c\kv}-E_{v,\kv-\qv}$, i.e., out of the entire sum over bands, we keep only the term with $n=c$, $m=v$, $f(E_{n\kv}=0)$,  $f(E_{v\kv-\qv}=1)$.  This is called the ``Tamm-Dancoff approximation" and it is usually  good when the system under investigation has a band gap.  Thus, the approximate form of $L_0$ for the two chosen bands is
%\vout{A major simplification can be achieved if we restrict to a static kernel. This is also called single pole approximation where the full excitonic two body wavefunctions are not frequency dependent.
%In the following derivation, we, firstly, discuss the BSE in the case of only resonant part ("Tamm-Dancoff approximation") is retained. This is usually a good approximation when the system under investigation has a band gap. Nevertheless, in the general case, the antiresonant terms have to be taken into account. For this reason, at the end of this section, we rewrite the e-h effective Hamiltonian in such general case.
%The full non-interacting electron-hole DFT Green's function can be rewritten in the one-electron wave functions are:}
\be\label{PPEH}
L_{0,(\alpha\beta,\gamma\delta)}(\kv,\pv,\qv;\omega) = i\delta_{\kv\pv}{\psi^{\alpha*}_{c\kv}\psi^{\gamma}_{c\kv}\frac{1}{\omega-E_{c\kv}+E_{v\kv-\qv}\pm i0^+}\psi^{\delta*}_{v\kv-\qv}\psi^{\beta}_{v\kv-\qv}} \,,
\ee
or, switching to the Bloch band representation
\be
L_{0,(cv,cv)}(\kv,\pv,\qv;\omega) \equiv i \frac{\delta_{\kv,\pv}}{\omega-E_{c\kv}+E_{v\kv-\qv}\pm i0^+}\,.
\ee
  
 The interacting electron-hole Green's function, restricted to the chosen pair of bands is then given by
 \be
   [L_{cv,cv}]^{-1}(\kv,\pv,\qv;\omega)= [L_{0,(cv,cv)}]^{-1}(\kv,\pv,\qv;\omega)-I_{cv,cv}(\kv,\pv,\qv)\,.
   \label{exc_ham}
 \ee
  where $I_{cv,cv}=I^x_{cv,cv}+I^d_{cv,cv}$ and 
  \be
I^{x}_{cv,cv}(\kv,\pv,\qv)=-i\sum_{\mu\nu}\sum_{\alpha\beta\gamma\delta}\psi^{\alpha*}_{c\kv}\psi^{\beta}_{v\kv-\qv}\Gamma^\mu_{\alpha\beta}(\kv,\kv-\qv)v_{\mu\nu}(\qv)\Gamma^\nu_{\delta\gamma}(\pv-\qv,\pv) \psi^{\delta^*}_{v\pv-\qv}\psi^{\gamma}_{c\pv}
\ee
and
\be
I^{d}_{cv,cv}(\kv,\pv,\qv)=\sum_{\mu\nu}\sum_{\alpha\beta\gamma\delta}\psi^{\alpha*}_{c\kv} \psi^{\gamma}_{c\kv}\Gamma^\mu_{\alpha\gamma}(\kv,\pv)w_{\mu\nu}(\kv-\pv)\Gamma^\nu_{\delta\beta}(\kv-\qv,\pv-\qv)\psi^{\delta*}_{v\pv-\qv}\psi^{\beta}_{v\pv-\qv}\,.
\ee
Finally, the matrix $ [L_{cv,cv}]^{-1}(\kv,\pv,\qv;\omega)$ for given momentum $\qv$ of the center of mass of the electron-hole system is identified as $\omega \delta_{\kv,\pv} - H^{2p,res}_{cv,cv}(\kv,\pv,\qv)$ where   
\be
H^{2p,res}_{cv,cv}(\kv,\pv,\qv) =  [E_{c\kv} - E_{v\kv-\qv}]\delta_{\kv,\pv}+ I_{cv,cv}(\kv,\pv,\qv)
\label{exc_kernel}
\ee
is the effective Hamiltonian matrix, with indices $\kv$ and $\pv$. Diagonalizing this Hamiltonian is equivalent to solving the Schr\"odinger equation for the relative motion of the electron and the hole, and gives the energy of the exciton.   The excitonic eigenvalues $\Omega^S_{cv}$ and eigenfunctions $A^{S}_{vc}(\kv)$ are obtained from the solution of the eigenvalue problem
  \begin{equation}
\sum_\pv H^{2p,res}_{vc,vc}(\kv,\pv,\qv) A^{S}_{vc,\qv}(\pv)=\Omega^S_{cv,\qv}A^{S}_{vc,\qv}(\kv)\,,
    \label{secular_eq.} 
   \end{equation}
where $\qv$ is the momentum of the center of mass of the exciton.
%The $H^{2p,res}_{vc,vc}(\kv,\kv,\qv\to 0)$  is usually diagonalized. This provides the excitonic eigenvalues $\Omega^S$ and eigenvectors $A^{S}_{vc}(\kv)$.
%
%
%\begin{equation}
% H^{2p,res}_{vc,vc}(\kv,\kv,\qv\to 0) A^{S}_{vc}(\kv)=\Omega_{S}A^{S}_{vc}(\kv)
% \label{secular_eq.}
%\end{equation}

The imaginary part of the macroscopic dielectric function  is derived, taking into account the  eqs. (\ref{L_real_space}): 
\be
L(\xiv_1,\xiv_2,\xiv_3,\xiv_4;\omega) = \sum_{\kv,\pv,\qv}\psi_{c\kv}(\xiv_1)[\psi_{v(\kv-\qv)}(\xiv_2)]^*L_{cv,cv}(\kv,\pv,\qv;\omega)[\psi_{c\pv}(\xiv_3)]^*\psi_{v,\pv-\qv}(\xiv_4)
\label{L_real_Space}
\ee
In the macroscopic limit, we have  $\pv=\kv$. Expanding Eq.~(\ref{P00}) with the help of Eqs.~(\ref{exc_kernel}), (\ref{L_real_Space}), and writing  $L_{cv,cv}=\sum_S A^{*S}_{vc,\qv}(\kv)\left(\omega-\Omega^S_{cv,\qv}\right)^{-1}A^{S}_{vc,\qv}(\kv) $ in the eigenvector basis set, we find

\begin{equation}
     - \Im m~\chi_P(\omega)= \frac{1}{N_c}\lim_{\qv \to 0}\frac{e^2}{q^2}\sum_S\sum_{{\bf k} vc}\left| \int d \xiv  \psi_{c,{\bf k}}(\xiv) e^{i\qv\cdot\rv} \psi^{*}_{v,{\bf k}}(\xiv)\right |^2 \left|A^{S}_{cv,\qv}(\kv) \right|^2   \pi\delta(\omega -\Omega^S_{cv\qv})\,,
      \label{eps_bse}
 \end{equation}
which is the generalization of the equation in ref. (\onlinecite{PhysRevB.62.4927}) to the quasi-2D case.

The above formulas take into account only the resonant part of the electron-hole propagator between the two selected bands.  When the antiresonant part is also considered, the e-h effective Hamiltonian $H^{2p,exc}$ becomes a $2\times 2$ matrix:

\be
 \left( \begin{array}{cc}
 H^{2p,res}_{vc,vc}&  I_{vc,cv}  \\
-(I_{vc,cv})^{*} &-(H^{2p,res}_{vc,vc})^* \\
 \end{array} \right)\,,
\ee
where all the entries are matrices with indices $\kv$ and $\pv$.  The  excitonic eigenvalues and eigenvectors are obtained by diagonalizing this more general Hamiltonian, resulting in  Casida-like equations \cite{casida_Eq}. 

\section{Summary and outlook}\label{summary}
In this paper we have proposed that calculations of the electronic structure of quasi-two dimensional materials be performed on a native two-dimensional basis, consisting of two-dimensional Bloch sums of numerically generated atomic orbitals, such as the ones used in the AIMS-FHI approach.  This approach removes the unphysical assumption of periodicity in the third dimension, which may be a cause of spurious results.    The formal structure that we have developed in this paper lays the foundation  for a direct two-dimensional numerical implementation of Kohn-Sham, GW, and Bethe-Salpeter equations in quasi 2D systems. We expect that the two-dimensional implementation of these tasks will speed up the calculations and make them more reliable, once the required computer codes are fully developed.  While it may take a considerable initial effort  to write the computer codes that will accomplish these tasks, we are confident that the payoff will be very high in the long run, resulting in a powerful computational tool occupying an intermediate position between those of molecular quantum chemistry and those of conventional solid state physics.

\section{Acknowledgements}  
  The authors thank A. H. Castro-Neto for his suggestions and support.
  PET gratefully acknowledges support from NRF under Grant No. R-723-000-001-281.
  GV gratefully acknowledges the hospitality of the Center for Advanced 2D Materials at NUS and support from DOE Grant No. DE-FG02-05ER46203.
 
   \appendix
   \section{GW: the Plasmon Pole Approximation (PPA)}\label{app_a}
 
    To simplify the self-energy calculations in ref. \onlinecite{PhysRevLett.55.1418}  a generalized plasmon pole approximation (PPA) was proposed in refs. \cite{PhysRevB.48.17791,PhysRevB.47.15404}.
    The main goal is to determine the inverse dielectric function $\epsilon^{-1}_{\alpha\beta}$
    \be
                \tilde \epsilon^{-1}_{\alpha\beta}(\qv,\omega) =S_{\alpha\beta}(\qv) +\left[ v^{1/2}(\qv)\right]_{\alpha\gamma}\Pi^{\gamma\delta}(\qv,\omega)\left[ v^{1/2}(\qv)\right]_{\delta\beta} 
                \label{diel_func_symm_dyn}
    \ee
   where $\Pi_{\alpha\beta}$ is the response function. This is related to the RPA Polarization function $P^{(0)}$ through:
   \be
	   \Pi^{-1}_{\alpha\beta}(\qv,\omega)=  \left( P^{(0)}\right)^{-1}_{\alpha\beta}(\qv,\omega) +v_{\alpha\beta}(\qv)
   \ee
   
   Diagonalizing the {\it static} Response Function:
   \be
   	   \Pi^{-1}_{\alpha\beta}(\qv,\omega=0)= \sum_l \Phi^l_{\alpha}(\qv)\bar \omega_l(\qv)\left( \Phi^l_{\beta}(\qv)\right)^*
   \ee
     with eigenvalues  $\bar \omega_l(\qv)$ (if positive just below the real axis in the complex plane and vice-versa if negative) and the eigenvectors $\Phi^l_{\alpha}$. Index $l$ runs over all spectrum. The eigenvectors $\Phi^l_{\alpha}(\qv)$ are related to the one electron wave function trough an unitary transformation $U$ $\Phi^l_{\alpha}(\qv)=U^{l}_{\alpha\beta}\psi^{\beta}_{\kv}$.
      In order to extend the dielectric function $\tilde \epsilon_{\alpha\beta}$ to the dynamic case, we only consider the $\omega$ dependence of the eigenvalues:
   we assume that the eigenvectors are not depending on the frequency (Plasmon Pole Model Approximation):
   So we have:
   \be
      	   \Pi^{-1}_{\alpha\beta}(\qv,\omega)= \sum_l \Phi^l_{\alpha}(\qv)\left( \omega -\bar \omega_l(\qv)\right) \left( \Phi^l_{\beta}(\qv)\right)^*
   \ee     
 
  From the eq.~(\ref{diel_func_symm_dyn}), the inverse of dielectric function can be written then:
    \be
               \tilde \epsilon_{\alpha\beta}^{-1}(\qv,\omega) = 1+ \sum_l \Phi^l_{\alpha}(\qv)\frac{Z_l(\qv)}{\omega-\bar \omega_l(\qv)}\left( \Phi^l_{\beta}(\qv)\right)^*           
    \ee   
   The renormalization factor $Z_l(\qv)$ is determined by the Johnson's generalized f-sum rule \cite{PhysRevB.9.4475}:
   
   \be
          \int_{0}^{\infty} d\omega \omega \Im m  \tilde \epsilon^{-1}_{\alpha\beta}(\qv,\omega)=-\frac{\pi}{2}\omega^{2}_{p}\frac{\rho_{\alpha\beta}}{\bar{\rho}}\,,
   \ee
   where $\omega^{2}_{p}$ is the plasma frequency, $\rho_{\alpha\beta}\equiv\sum_{n\kv} \psi_{n\kv}^{\alpha}\left( \psi_{n\kv}^{\beta}\right)^{*}$ the density matrix and $\bar \rho\equiv\sum_{\alpha} \rho_{\alpha\alpha}$ is the density mean value.
\section{GW with no plasmon-pole model}\label{app_b}
\begin{figure}
\includegraphics[width=4.0 in]{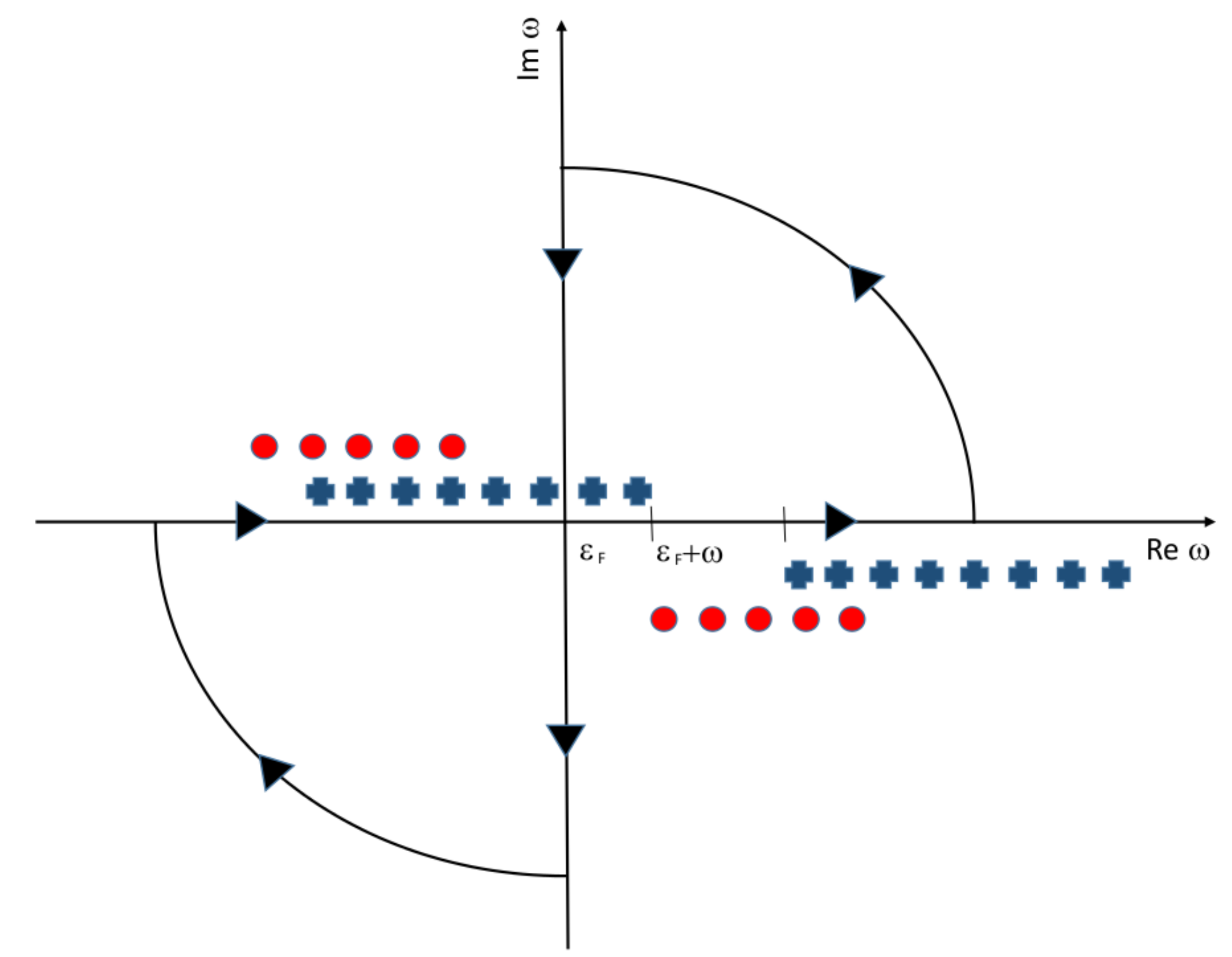}
\caption{Contour integration in the $\omega$ complex plane. The blue crosses stand for the Green function poles whereas the red balls represent the Screened potential $W_{c}$ poles.}
\label{contour}
\end{figure}

In this section, we explicit the calculation of the Self energy frequency integral. We follow the original work of Leb\`esgue et al. \cite{Lebesgue_prb} called {\it Contour integral method}. The idea is to avoid the real axis of frequency where the poles of the $G$ and the $W$ lie by using a Matsubara rotation and a closed contour as shown in Fig. (\ref{contour}). 
In order to explicit the integration is worth to dividing the screened interaction in the {\it bare} Coulomb potential and  {\it correlation} part: $W\equiv v+W_{c}$.
This division brings to split the self energy in two parts $\Sigma\equiv\Sigma^x+\Sigma^c $, respectively the {\it exchange} and {\it correlation} terms.
The exchange $\Sigma^x$ is determined by the Coulomb potential and after frequency integration:

\be
\begin{split}
\Sigma^{x}_{\mu\nu}(\pv) &=\frac{i}{2\pi}\sum_{\qv}\int d\omega e^{-i\eta\omega}G_{0}^{\alpha\beta}(\pv-\qv,\omega)\Gamma^{\delta}_{\beta\nu}(\pv-\qv,\pv) v_{\gamma\delta}(\qv)\\
&=
-\sum_{\qv,n}\Theta(\epsilon_{F}-E_{n(\pv-\qv)})\Gamma^{\gamma}_{\mu\alpha}(\pv,\pv-\qv)\psi^{\alpha}_{n(\pv-\qv)}\psi^{\beta*}_{n(\pv-\qv)}\Gamma^{\delta}_{\beta\nu}(\pv-\qv,\pv) v_{\gamma\delta}(\qv)
\end{split}
\ee

In order to determine the correlation Self Energy $\Sigma^{c}_{\mu\nu}(\pv,\omega)$ is customary to rewrite the frequency convolution integral in (omitting the momentum coordinates)

\be
\Sigma^{c}(\omega)=\int d\omega' e^{i\eta\omega'} G(\omega+\omega')W_{c}(\omega') 
\ee

This is allowed considering that $W_{c}(\omega)=W_{c}(-\omega)$.
The full contour integration performed on the path shown in Fig (\ref{contour}) is equal to the residues due to the poles contribution inside the contour. This is equal to the integral on the real axis plus the integral on the imaginary axis (the integral on the parts of the circle is zero since the Self energy has an advanced and retarded part). In particular, the $W_{c}$ poles are out of the contour whereas only some poles of the Green function are included inside.
Therefore, we have (taking advantage of $W_{c}(-i\omega)=W^*_{c}(i\omega)$) that integral on the $\omega$ imaginary axis:
 \be
 \int_{-\infty}^{-\infty} d (i\omega') \frac{\omega-\epsilon_n-i\omega'}{(\omega-\epsilon_n)^2 +\omega'}W_c(i\omega')=2i \int_{0}^{\infty}d (\omega') \frac{\omega-\epsilon_n}{(\omega-\epsilon_n)^2 +\omega'^2}\Re W_c(i\omega')
 \ee
The final form of the Self Energy integration is:

\be
\begin{split}
&\Sigma^{c}_{\mu\nu}(\pv)=-\sum_{\qv,n}\Gamma^{\gamma}_{\mu\alpha}(\pv,\pv-\qv)\psi^{\alpha}_{n(\pv-\qv)}\psi^{\beta*}_{n(\pv-\qv)}\Gamma^{\delta}_{\beta\nu}(\pv-\qv,\pv)*  \\
 &\left[  \frac{1}{\pi} \int_{0}^{\infty}d \omega' \frac{\omega-\epsilon_n}{(\omega-\epsilon_n)^2 +\omega'^2}\Re W_c(\qv,i\omega')+\left( \theta(\epsilon_F-\epsilon_n)\theta(\epsilon_n-\omega) -\theta(\epsilon_n-\epsilon_F)\theta(\omega-\epsilon_n)\right) W_c(\qv,|\epsilon_n-\omega|) \right]
\end{split}
 \ee

  \section{Numeric atom-centered orbitals in FHI-AIMS}\label{app_c}
  
 In this appendix, we describe the generation of NAOs in FHI-AIMS as described in ref.\cite{Fhi-Aims}. In eq. (\ref{CovariantBasis}), the atomic orbital $\phi_{\alpha}$ are usually taken to be Gaussians. Alternatively in FHI-AIMS we have numerical atomic orbitals
   \be
	   \phi_{\alpha}(\xiv)= \frac{u_{\alpha}(\xi)}{\xi}Y_{lm}(\Omega)\,,
   \ee
   where $u_{\alpha}(\xi)$ is the radial shape, which is numerically tabulated, and $Y_{lm}$ are spherical functions of $\Omega=(\theta,\phi)$, and $\xiv=(\xi,\Omega)$.
   The $u_{\alpha}(\xi)$ are the numerical solutions of (possibly scalar relativistic) Schr\"odinger-like equations:
   \be\label{Schro_eq}
   \left[ -\frac{1}{2}\frac{d^2}{d\xi^2}-\frac{1}{\xi}\frac{d}{d\xi}+\frac{l(l+1)}{2\xi^2}+v_{\alpha}(\xi)+v_{cut}(\xi)\right]u_{\alpha}(\xi)=\varepsilon_{\alpha} u_{\alpha}(\xi)
   \ee
    with $\varepsilon_{\alpha}$ the eigenvalues.
   The generating potential for the radial wave functions  consists of two parts -- an atomic-like part  $v_{\alpha} (\xi)$ and a steeply increasing confining potential $v_{cut}(\xi)$, which takes into account the atomic environment. The practical numerical implementation of the solution of Eq. (\ref{Schro_eq}) is described in ref.  \onlinecite{Fhi-Aims}.
   In brief, the minimal basis consists of the core and valence functions of spherically symmetric free atoms. The $v_{\alpha} (\xi)$ is set to the self-consistent free-atom radial potential.
   The confining potential $v_{cut}(\xi)$ is defined by a smooth analytical shape: 
 \begin{equation}
   v_{cut}(\xi)=\begin{cases}
     0 &  \xi\le \xi_{onset}\\
     s\frac{1}{(\xi-\xi_{cut})^2}\cdot \exp \left(\frac{w}{\xi-\xi_{onset}}\right) & \xi_{onset} <\xi<\xi_{cut}\\
     \infty & \xi\ge \xi_{cut}
   \end{cases}
 \end{equation}
 with the width $w=\xi_{onset}-\xi_{cut}$ and $s$ a scaling parameter.
 The $v_{cut}$ is, therefore, determined in such a way that smoothly decreases up to $r_{cut}$ confinement.

%  \begin{thebibliography}{[Vo]}
%
%
%\bibliographystyle{unsrt}

%\bibliography{2dbib}{}
 %\begin{thebibliography}{}

\end{document}